\documentclass[12pt]{amsart}
\usepackage{amscd}
\usepackage{amsfonts}
\usepackage{epsf}
\usepackage{epic}
\usepackage{eepic}
\usepackage{latexsym}

\begin{document}

\title[Chern-Simons Invariants Of Closed Hyperbolic 3-Manifolds]
{Chern-Simons Invariants Of Closed Hyperbolic 3-Manifolds}  
\author{A.A. Bytsenko}
\address{Departamento de Fisica, Universidade Estadual de Londrina,
Caixa Postal 6001, Londrina-Parana, Brazil; on leave from Sankt-Petersburg
State Technical University, Russia \,\, {\em E-mail address:} 
{\rm abyts@fisica.uel.br}}
\author{A.E. Gon\c calves}
\address{Departamento de Fisica, Universidade Estadual de Londrina,
Caixa Postal 6001, Londrina-Parana, Brazil\,\, {\em E-mail address:} 
{\rm goncalve@fisica.uel.br}}
\author{F.L. Williams}
\address{Department of Mathematics, University of Massachusetts,
Amherst, Massachusetts 01003, USA\,\, {\em E-mail address:} 
{\rm williams@math.umass.edu}}

\date{May, 1999}

\thanks{First author partially supported by a CNPq grant (Brazil), RFFI 
grant (Russia) No 98-02-18380-a, and by GRACENAS grant (Russia) No 6-18-1997.}

\maketitle

\begin{abstract}

The Chern-Simons invariants of irreducible $U(n)-$ flat connections on 
compact hyperbolic 3-manifolds of the form $\Gamma\backslash {\Bbb H}^3$
are derived. The explicit formula for the Chern-Simons functional is given
in terms of Selberg type zeta functions related to the twisted eta
invariants of Atiyah-Patodi-Singer.

\end{abstract}
\vspace{0.3cm}
\vspace{0.3cm}

\section{Introduction}

A gauge Chern-Simons theory is interesting both for its mathematical
novelty and for its applications for certain planar condensed matter 
phenomena (such as the fractional quantum Hall effect) and for nonabelian 
gauge models in field theory. For example, the supersymmetric 
index can be computed in ${\mathcal  N}=1$ super
Yang-Mills theory in three spacetime dimensions with a Chern-Simons
interaction at level $k$ by making a relation to a Chern-Simons theory 
\cite{witt99}. 

Recently the Chern-Simons functional has been actively 
studied in low dimensional 
topology in connection with Floer homology \cite{floe88-118}, where 
it was used as a Morse function. This functional has also been used in 
topological field theory, where the new 
topological invariants of oriented 3-manifolds and links in it were
specified \cite{witt89-121}. The topology of manifolds can be studied with
the help of the path integral approach in quantum field theory, which
suggests the asymptotic behaviour of invariants including the
Chern-Simons invariant, eta invariant, Redemeister torsion, and so on
\cite{moor89-123,dijk90-129,kirb91-105,free91-141,jeff92-147,roza96-175,
adam95,byts97-505,byts98-13}.
Under such circumstances it is important to have an explicit formula for
the Chern-Simons invariant.

The Chern-Simons invariants for the case of $SU(2)$ and for some classes of 
3-manifolds (including the Seifert fibered 3-manifolds) have been listed in 
\cite{kirk90-287,kirk93-153,auck94-115} by cutting a 3-manifold into pieces 
for which the invariants can be computed. The formula for the Chern-Simons
invariants of irreducible $SU(n)-$flat connection on the Seifert fibered
3-manifolds has been derived in \cite{nish98-9}.
In this paper we shall compute this invariant using the index theorem for 
a manifold with a closed hyperbolic boundary 
$X_{\Gamma}=\Gamma\backslash {\Bbb H}^3$.

\section{The Chern-Simons invariant}

In this section we briefly summarize the formalism that we shall use in
this paper. Our goal is to present a formula for the $U(n)$-
Chern-Simons invariant of an irreducible flat connection on a closed
hyperbolic 3-manifold. This invariant is defined by the Chern-Simons
functional $CS$ which can be considered as a function on a space of
connections on a trivial principal bundle over a compact oriented
3-manifold $X_{\Gamma}$ given by

$$
CS(A)=\frac{1}{8\pi^2}\int_{X_{\Gamma}}\mbox{Tr}\left(A\wedge dA+\frac{2}{3}A
\wedge A\wedge A\right)\mbox{.}
\eqno{(2.1)}
$$
Let $X$ be a locally symmetric Riemannian manifold with negative sectional 
curvature. Its universal covering ${\widetilde   X}\rightarrow X$ is a 
Riemannian symmetric space of rank one.
The group of orientation preserving isometries ${\widetilde G}$ of 
${\widetilde   X}$ is a
connected semisimple Lie group of real rank one and 
${\widetilde   X}={\widetilde   G}/{\widetilde   K}$,
where ${\widetilde   K}$ is a maximal compact subgroup of ${\widetilde   G}$. 
The fundamental group of $X$ acts by covering transformations on 
${\widetilde   X}$ and gives
rise to a discrete, co-compact subgroup $\Gamma \subset {\widetilde   G}$ 
such that $X=\Gamma\backslash {\widetilde   G}/{\widetilde   K}$. 
If $G$ is a linear connected finite 
covering of ${\widetilde   G}$, the embedding 
$\Gamma\hookrightarrow {\widetilde   G}$ lifts
to an embedding $\Gamma\hookrightarrow G$. Let $K\subset G$ be a maximal
compact subgroup of $G$, then $X_{\Gamma}=\Gamma\backslash G/K$ is a compact
manifold.

Let ${\rm g}={\frak k}\oplus {\frak p}$ be a
Cartan decomposition of the Lie algebra ${\rm g}$ of $G$. Let
${\frak a}\subset{\frak p}$ be a 
one-dimensional subspace and let $J=K\bigcap G_{{\frak a}}$ be the 
centralizer of ${\frak a}$ in $K$. Fixing a positive root system of 
$({\rm g},{\frak a})$ we have an Iwasawa decomposition 
${\rm g}={\frak k}\oplus {\frak a}\oplus {\frak n}$. For
$G=SO(m,1)$ \,$(m\in {\Bbb Z}_{+})$, $K=SO(m)$, and $J=SO(m-1)$. 
The corresponding symmetric space of non-compact type is the real hyperbolic
space ${\Bbb H}^m$ of sectional curvature $-1$. It's compact dual space is
the unit $m-$ sphere.

Let ${\Bbb P}=X_{\Gamma}\otimes {\frak G}$ be a trivial principal bundle 
over $X_{\Gamma}$ with
the gauge group ${\frak G}=U(n)$ and let 
${\frak A}_{X_{\Gamma}}=\Omega^1(X_{\Gamma},{\frak g})$ be the
space of all connections on ${\Bbb P}$; this space is an affine space of 
1-forms on $X_{\Gamma}$ with values in the Lie algebra ${\frak g}$ of 
${\frak G}$.
The value of the function $CS(A)$ on the space of connections 
${\frak A}_{X_{\Gamma}},\, A\in{\frak A}_{X_{\Gamma}}$, at a critical point 
can be regarded as a topological invariant of a pair $(X_{\Gamma},\chi)$, 
where $\chi$ is an orthogonal 
representation of the fundamental group $\pi_1(X_{\Gamma})\equiv \Gamma$. 

Let ${\frak A}_{X_{\Gamma},F}=\{A\in{\frak A}_{X_{\Gamma}}|F_A= 
dA+A\wedge A=0\}$
be the space of flat connections on ${\Bbb P}$. An important formula 
related to the integrand in (2.1) is

$$
d{\rm Tr}\left(A\wedge dA+\frac{2}{3}A\wedge A\wedge A\right)
={\rm Tr}\left(F_{A}\wedge F_{A}\right)
\mbox{.}
\eqno{(2.2)}
$$

Eq. (2.2) gives another approach to the Chern-Simons invariant. 
Indeed, let $M$ be an oriented 4-manifold with boundary 
$\partial M=X_{\Gamma}$.
One can extend ${\Bbb P}$ to a trivial ${\frak G}-$ bundle over $M$; then  
Stokes' theorem gives

$$
CS(A)=\frac{1}{8\pi^2}\int_{M}\mbox{Tr}\left(F_{{\widetilde   A}}\wedge 
F_{{\widetilde   A}}\right)
\mbox{,}
\eqno{(2.3)}
$$
where ${\widetilde   A}$ is any extension of $A$ over $M$. 
Eq. (2.3) can be viewed as
a generalization of the Chern-Simons invariant to the case in which $\Bbb P$ 
is a non-trivial $U(n)-$ bundle over $X_{\Gamma}$ \cite{nish98-9}. In fact 
we shall show that the Chern-Simons functional is well-defined 
${\rm modulo} ({\Bbb Z}/2)$ in the case of a $U(n)-$ connection.

\section{Invariants of closed hyperbolic manifolds}

Let ${\Bbb E}_{\chi}={\widetilde   X}\otimes_{\chi}{\Bbb C}^n$ be a flat 
vector bundle, and let ${\widetilde  {\Bbb P}}$ be a principal $U(n)-$ 
bundle over 
$X_{\Gamma}$ which is an extension of $\Bbb P$. Any 1-dimensional 
representation $\chi$ of
$\Gamma$ factors through a representation of $H^1(X_{\Gamma},{\Bbb Z})$. 
It can be
shown that for a unitary representation $\chi: \Gamma\rightarrow U(n)$,
the corresponding flat vector bundle ${\Bbb E}_{\chi}$ is topologically trivial
$({\Bbb E}_{\chi}\cong X_{\Gamma}\otimes {\Bbb C}^n)$ iff 
${\rm det}\,\chi|_{{\rm Tor}^1}: {\rm Tor}^1\rightarrow U(1)$ is the trivial 
representation.
Here ${\rm Tor}^1$ is the torsion part of $H^1(X_{\Gamma},{\Bbb Z})$ and 
${\rm det}\,\chi$ is a 1-dimensional representation of $\Gamma$ defined by
${\rm det}\,\chi(\gamma):={\rm det}\,(\chi(\gamma))$, for $\gamma\in \Gamma$.

For any representation $\chi: \Gamma\rightarrow U(n)$ one can construct a 
vector bundle ${\Bbb {\widetilde   E}}_{\chi}$ over a certain 4-manifold 
$M$ with 
boundary $\partial M=X_{\Gamma}=\Gamma\backslash{\Bbb H}^3$ which is an 
extension of a 
flat vector bundle ${\Bbb E}_{\chi}$ over $\Gamma\backslash{\Bbb H}^3$.
Let ${\widetilde   A}_{\chi}$ be any extension of a flat connection 
$A_{\chi}$
corresponding to $\chi$. The index theorem of Atiyah-Patodi-Singer for the 
twisted Dirac operator $D_{{\widetilde   A}_{\chi}}$ 
\cite{atiy75-77,atiy75-78,atiy76-79} is given by

$$
{\rm Ind}\left(D_{{\widetilde A}_{\chi}}\right)=\int_{M}{\rm ch}
({\Bbb {\widetilde   E}}_{\chi})
{\widehat  A}(M)-\frac{1}{2}(\eta(0,{\mathcal  O}_{\chi})+
h(0,{\mathcal  O}_{\chi}))
\mbox{,}
\eqno{(3.1)}
$$
where $h(0,{\mathcal  O}_{\chi})$ is the dimension of the space of harmonic 
spinors on $X_{\Gamma}$ ($h(0,{\mathcal  O}_{\chi})
={\rm dim}{\rm ker}\,{\mathcal  O}_{\chi}$ = 
multiplicity of the 0-eigenvalue of ${\mathcal  O}_{\chi}$ acting on 
$X_{\Gamma}$); ${\mathcal O}_{\chi}$ is a Dirac operator on 
$X_{\Gamma}$ acting on spinors with coefficients in $\chi$ (for details see 
Appendices A and B). 

The Chern-Simons invariants of $\Gamma\backslash{\Bbb H}^3$ can be derived
from Eq. (3.1) as follows. For a closed manifold $M$ we have

$$
-\frac{1}{8\pi^2}{\rm Tr}\left(F_{{\widetilde   A}_{\chi}}\wedge
F_{{\widetilde   A}_{\chi}}\right) ={\rm ch}_2({\Bbb {\widetilde   E}}_{\chi}) 
\mbox{,}
\eqno{(3.2)}
$$
where ${\rm ch}_2({\Bbb {\widetilde   E}}_{\chi})$ is the second Chern 
character of ${\Bbb {\widetilde   E}}_{\chi}$, which is expressed in terms 
of the first and second Chern classes:

$$
{\rm ch}_2({\Bbb {\widetilde   E}}_{\chi})=
\frac{1}{2}c_1({\Bbb {\widetilde E}}_{\chi})^2-
c_2({\Bbb {\widetilde   E}}_{\chi})
\mbox{.}
\eqno{(3.3)}
$$
The Chern character and the $\widehat  A-$ genus are given by 

$$
{\rm ch}({\Bbb {\widetilde   E}}_{\chi})={\rm rank}\,
{\Bbb {\widetilde   E}}_{\chi}+
c_1({\Bbb {\widetilde   E}}_{\chi})
+{\rm ch}_2({\Bbb {\widetilde   E}}_{\chi})
$$
$$
={\rm dim}\,\chi +c_1({\Bbb {\widetilde   E}}_{\chi})+
{\rm ch}_2({\Bbb {\widetilde   E}}_{\chi})
\mbox{,}
\eqno{(3.4)}
$$
and
$$
{\widehat  A}(M)=1-\frac{1}{24}p_1(M)
\mbox{.}
\eqno{(3.5)}
$$
Thus we have 

$$
{\rm ch}({\Bbb {\widetilde   E}}_{\chi}){\widehat  A}(M)=\left({\rm dim}\,\chi+
c_1({\Bbb {\widetilde   E}}_{\chi})+
{\rm ch}_2({\Bbb {\widetilde  E}}_{\chi})\right)
\left(1-\frac{1}{24}p_1(M)\right)
$$
$$
={\rm dim}\,\chi+c_1({\Bbb {\widetilde   E}}_{\chi})+
{\rm ch}_2({\Bbb {\widetilde   E}}_{\chi})-
\frac{{\rm dim}\,\chi}{24}p_1(M)
\mbox{.}
\eqno{(3.6)}
$$
The integral over the four manifold $M$ takes the form

$$
\int_M {\rm ch}({\Bbb {\widetilde   E}}_{\chi}){\widehat  A}(M)=
\int_M {\rm ch}_2({\Bbb {\widetilde   E}}_{\chi})-\frac{{\rm dim}\,
\chi}{24}\int_M p_1(M)
$$
$$
=-\frac{1}{8\pi^2}\int_M {\rm Tr}\left(F_{{\widetilde   A}_{\chi}}\wedge
F_{{\widetilde   A}_{\chi}}\right)-\frac{{\rm dim}\,\chi}{24}\int_M p_1(M)
\mbox{,}
\eqno{(3.7)}
$$
and we have

$$
{\rm Ind}\left(D_{{\widetilde   A}_{\chi}}\right)=-\frac{1}{8\pi^2}\int_M
{\rm Tr}\left(F_{{\widetilde   A}_{\chi}}\wedge
F_{{\widetilde   A}_{\chi}}\right)
$$
$$
-\frac{{\rm dim}\,\chi}{24}\int_M p_1(M)
-\frac{1}{2}(\eta(0, {\mathcal  O}_{\chi})+h(0,{\mathcal  O}_{\chi}))
\mbox{.}
\eqno{(3.8)}
$$
For a trivial representation $\chi_0$ one can take a trivial flat connection
${\widetilde   A}\equiv{\widetilde   A}_{{\chi}_0}$; 
then $F_{{\widetilde   A}_{{\chi}_0}}=0$ and 
for this choice we get

$$
{\rm Ind}\left(D_{{\widetilde A}_{\chi_0}}\right)=-\frac{1}{24}
\int_M p_1(M)-\frac{1}{2}\left(\eta(0,{\mathcal  O})+h(0,{\mathcal  O})\right)
\mbox{.}
\eqno{(3.9)}
$$
Using Eqs. (3.8) and (3.9) one can obtain  

$$
{\rm Ind}\left(D_{{\widetilde A}_{\chi}}\right)-{\rm dim}\chi
{\rm Ind}\left(D_{{\widetilde A}_{\chi_0}}\right)
=-\frac{1}{8\pi^2}\int_M
{\rm Tr}\left(F_{{\widetilde   A}_{\chi}}\wedge
F_{{\widetilde   A}_{\chi}}\right)
$$
$$
-\frac{1}{2}\left(\eta(0,{\mathcal  O}_{\chi})-
{\rm dim}\,\chi\eta(0,{\mathcal  O})\right)
\,\,\,\,\,\,\,\,\,\, {\rm modulo}({\Bbb Z}/2)
\mbox{,}
\eqno{(3.10)}
$$
and 

$$
CS(\chi)\equiv 
\frac{1}{2}\left({\rm dim}{\chi}\eta(0,{\mathcal  O})-
\eta(0,{\mathcal  O}_{\chi})\right)
\,\,\,\,\,\,\,\,\,\, {\rm modulo}({\Bbb Z}/2)
\mbox{.}
\eqno{(3.11)}
$$
We are now able to use the formulae (C.3) and (C.6) of Appendix C for the eta 
invariant in Eq. (3.11).
It follows that

$$
Z(0,{\mathcal  O}_{\chi})=Z(0,{\mathcal  O})^{{\rm dim}\,{\chi}}
e^{-2\pi i CS(\chi)}
\mbox{.}
\eqno{(3.12)}
$$
Finally the Chern-Simons functional takes the form

$$
CS(\chi)=\frac{1}{2\pi i}{\rm log}\left[
\frac{Z(0,{\mathcal  O})^{{\rm dim}\,\chi}}{Z(0,{\mathcal  O}_{\chi})}\right]
\,\,\,\,\,\,\,\,\,\, {\rm modulo}({\Bbb Z}/2)
\mbox{.}
\eqno{(3.13)}
$$

\section{Concluding remarks}

For an irreducible 
$U(n)-$ representation formula (3.13) gives the values of the Chern-Simons 
invariants. This is the main result of the paper. The formula for the
Chern-Simons functional involves a Selberg type zeta function 
$Z(0,{\mathcal  O}_{\chi})$ associated with
twisted eta invariants of Atiyah-Patodi-Singer 
\cite{atiy75-77,mill78-108,mosc}.
In the case of a $U(n)-$ connection this is a well-defined functional, 
${\rm modulo}({\Bbb Z}/2)$. Note also that the explicit results
(3.12) and (3.13) can be very important for the study of the relation between 
quantum invariant
for an oriented 3-manifold, defined with the help of a representation theory 
of quantum group \cite{resh91-103}, and  Witten's invariant \cite{witt89-121},
related to the path integral approach. We hope to return to this study
elsewhere.

\section{Appendix A. Dirac operators and the eta invariant}

We recall in this section some standard material on Dirac bundles
(for details see, for example, Ref. \cite{laws89}). Let $M$ be an 
even-dimensional orientable Riemannian manifold with boundary
$\partial M\equiv Y$ and let ${\Bbb E}\rightarrow M$ be a hermitian 
vector bundle equipped
with a compatible connection $\nabla$. Let ${\Bbb C}\ell(M)$ denote 
the complexified Clifford bundle. We suppose that there is a bundle map from
${\Bbb C}\ell(M)\rightarrow {\rm End}\,{\Bbb E}$ which is an algebra 
homomorphism on each fiber and which covers the identity:
\newpage
\vspace{1cm}
$$
\begin{picture}(120,80)
\put(60,30){$M$}
\put(0,80){${\Bbb C}\ell(M)$}
\put(100,80){${\rm End}\,{\Bbb E}$}
\put(40,83){\vector(1,0){52}}
\put(28,75){\vector(1,-1){30}}
\put(105,75){\vector(-1,-1){30}}
\put(210,50){$(A.1)$}
\end{picture}
$$
is commutative. Denote by ${\Bbb S}$ the spin bundles associated to the two
half-spin representation of ${\rm Spin}({\rm dim}M)$; then 
${\Bbb S}={\Bbb S}^{+}\oplus{\Bbb S}^{-}$ where ${\Bbb S}^{\pm}$ are 
half-spin bundles. $\nabla$ induces a connection on 
${\Bbb S}\otimes{\Bbb E}$ which is compatible with both the metric and the 
${\Bbb Z}_2-$ grading. The latter connection canonically defines a Dirac 
operator $D: C^{\infty}({\Bbb S}^{+}\otimes{\Bbb E})\rightarrow 
C^{\infty}({\Bbb S}^{-}\otimes{\Bbb E})$ described by

$$
D: C^{\infty}({\Bbb S}_{\Bbb E}^{+})\stackrel{\nabla}{\longrightarrow}
C^{\infty}(T^{*}M\otimes {\Bbb S}_{\Bbb E}^{+})\stackrel{{\frak M}}
{\longrightarrow}C^{\infty}({\Bbb S}_{\Bbb E}^{-})
\mbox{,}
\eqno{(A.2)}
$$
\vspace{0.3cm}

\noindent 
where ${\frak M}: T^{*}M\longrightarrow 
{\rm Hom}({\Bbb S}^{+}_{\Bbb E},
{\Bbb S}^{-}_{\Bbb E})$ denotes the Clifford multiplication and 
${\Bbb S}^{\pm}_{\Bbb E}\stackrel{def}{=}{\Bbb S}^{\pm}\otimes{\Bbb E}$.

Suppose the metric on a neighborhood ${\frak U}\simeq [0,1)\otimes Y$ of 
the boundary 
is a product metric. Then near the boundary $Y$ one has a separation of 
variables which leads to the representation 
$D=\sigma\left(\frac{\partial}{\partial x}+{\mathcal  O}\right)$, where the 
decomposition $\phi = dx\wedge \phi_1(x)+\phi_2(x)$ of a smooth form $\phi$
is used. Thus $D$ acts on 
$C^{\infty}\left((0,1),\Omega^p(Y)\oplus\Omega^p(Y)\right)$,
$\Omega^p(Y)$
being the space of smooth $p-$ forms on $Y$. A little computation shows
that ${\mathcal  O}$ is an elliptic self-adjoint map on
$\Omega^p(Y)$ defined by 
${\mathcal  O}\phi=(-1)^{k+p+1}(\varepsilon*d-d*)\phi$
\cite{atiy75-77}, where the symbol $*$ denotes 
the duality operator on $Y$, $k=({\rm dim}\,Y)/4$ and 
$\phi$ is either a
$2p-$ form $(\varepsilon=1)$ or a $(2p-1)-$ form $(\varepsilon=-1)$.
The operator ${\mathcal  O}$ maps even (odd) forms to even (odd) forms and can
therefore be decomposed as follows: 
${\mathcal  O}={\mathcal  O}^{ev}\oplus{\mathcal  O}^{odd}$, 
where ${\mathcal  O}^{ev}$ is isomorphic
to ${\mathcal  O}^{odd}$ \cite{atiy75-77}.

Recall some standard material on the eta invariant
of a self-adjoint elliptic differential operator acting on a compact 
manifold $Y$.
For details we refer the reader to Refs. \cite{atiy75-77,atiy75-78,atiy76-79} 
where the eta invariant
was introduced in connection with the index theorem for a manifold with
boundary. To define a spectral invariant which measures the asymmetry of the
spectrum ${\rm Spec}({\mathcal  O})$ of an operator ${\mathcal  O}$, 
one starts with next proposition.

\medskip
\par \noindent
{\bf Proposition A.1.}\,\,\,{\em The holomorphic function

$$
\eta(s,{\mathcal  O})\stackrel{def}{=}
\sum_{\lambda\in {\rm Spec}{\mathcal  O}\backslash\{0\}}
{\rm sgn}(\lambda)|\lambda|^{-s}={\rm Tr}\left({\mathcal  O}
\left({\mathcal  O}^2\right)^{-(s+1)/2}\right)
\mbox{,}
\eqno{(A.3)}
$$
is well defined for all $\Re s\gg 0$ and extends to a meromorphic function
on ${\Bbb C}$}.

\medskip

Indeed, from the asymptotic behaviour of the heat operator at $t=0$,
${\rm Tr}\left({\mathcal  O}\exp{(-t{\mathcal  O}^2)}\right)=O(t^{1/2})$ 
\cite{bism86-107} and from the identity

$$
\eta(s,{\mathcal  O})
=\frac{1}{\Gamma\left(\frac{s+1}{2}\right)}\int_{{\Bbb R}_{+}}
{\rm Tr}\left({\mathcal  O} e^{-t{\mathcal  O}^2}\right)t^{(s-1)/2}dt
\mbox{,}
\eqno{(A.4)}
$$
it follows that $\eta(s,{\mathcal  O})$ admits a meromorphic extension to the 
whole $s-$ plane, with at most simple poles at 
$s=({\rm dim}\,Y-k)/({\rm ord}\,{\mathcal  O})$\,\,\,$(k\in {\Bbb Z}_{+})$ and
locally computable residues.

It has been established that point $s=0$ is not a pole, which makes it
possible to define the eta invariant of ${\mathcal  O}$ by 
$\eta(0,{\mathcal  O})$.
It also follows directly that 
$\eta(0,-{\mathcal  O})=-\eta(0,{\mathcal  O})$ and
$\eta(0,\lambda{\mathcal  O})=\eta(0,{\mathcal  O})$, $\forall \lambda>0$. 
One can attach the eta invariant to any operator of Dirac type on a compact
Riemannian manifold of odd dimension. Dirac operators on even dimensional
manifolds have symmetric spectrums and, therefore, trivial eta invariants.
Because ${\mathcal  O}^{ev}$ is isomorphic to ${\mathcal  O}^{odd}$, we have

$$
\eta(s,{\mathcal  O}^{ev})=\eta(s,{\mathcal  O}^{odd})=
\frac{1}{2}\eta(s,{\mathcal  O})
\mbox{.}
\eqno{(A.5)}
$$

\section{Appendix B. Closed geodesics on negative curvature manifold and 
Selberg type zeta function}

Let $Y$ be a compact oriented $(4m-1)-$ dimensional Riemannian manifold of
constant negative curvature. A remarkable formula relating 
$\eta(s,B)$,\, $B\equiv{\mathcal  O}^{ev}$, to the closed geodesics on $Y$ 
has been derived by Millson \cite{mill78-108}. More explicitly, Millson 
proved the following result for a Selberg type (Shintani) zeta function.

\medskip
\par \noindent
{\bf Theorem B.1} (Millson [20]).\,\,\,{\em Define a zeta function by the 
following series which is absolutely convergent for $\Re\,s>0$

$$
{\rm log}Z(s,B)\stackrel{def}{=}\sum_{[\gamma]\neq 1}
\frac{{\rm Tr}\tau^{+}_{\gamma}-{\rm Tr}\tau^{-}_{\gamma}}
{|{\rm det}(I-P_h(\gamma))|^{1/2}}\frac{e^{-s\ell(\gamma)}}{m(\gamma)}
\mbox{,}
\eqno{(B.1)}
$$
where $[\gamma]$ runs over the nontrivial conjugacy classes in 
$\Gamma=\pi_1(Y)$, $\ell(\gamma)$ is the length of the closed geodesic 
$c_{\gamma}$ (with multiplicity $m(\gamma)$) in the free homotopy class
corresponding to $[\gamma]$, $P_h(\gamma)$ is the restriction of the linear
Poincar\'{e} map $P(\gamma)=d\Phi_1$ at $(c_{\gamma},\dot{c}_{\gamma})\in TY$
to the directions normal to the geodesic flow $\Phi_t$ and 
$\tau^{\pm}_{\gamma}$ is parallel translation around $c_{\gamma}$ on
$\Lambda^{\pm}_{\gamma}=\pm i$ eigenspace of 
$\sigma_B(\dot{c}_{\gamma})$ ($\sigma_B$ denoting the principal symbol of $B$).
Then $Z(s,B)$ admits a meromorphic continuation to the entire complex plane,
which in particular is holomorphic at $0$. Moreover

$$
{\rm log}Z(0,B)=\pi i\eta(0,B)
\mbox{.}
\eqno{(B.2)}
$$
}

\medskip

In fact $Z(s,B)$ satisfies the functional equation

$$
Z(s,B)Z(-s,B)=e^{2\pi i\eta(0,B)}
\mbox{.}
\eqno{(B.3)}
$$

\section{Appendix C. Twisted eta invariants associated to locally symmetric 
spaces}
 
Milson's formulae have been extended by Moscovici and Stanton \cite{mosc}
to Dirac type operators (acting in non-positively curved locally symmetric 
manifolds), even with additional coefficients in locally flat bundles. We 
now present the main results obtained in \cite{mosc}.

Let ${\widetilde   Y}$ denote be a simply connected cover of $Y$ which is 
a symmetric space of noncompact type, and let ${\Bbb {\widetilde   E}}$ 
denote the pull-back to $\widetilde   Y$ for any vector bundle
${\Bbb E}$ over $Y$. Bundles which will be considered satisfy a local 
homogeneity condition. Namely, a vector bundle ${\Bbb E}$ over $Y$ is 
${\mathcal  G}-$ locally homogeneous, for some Lie group ${\mathcal G}$, if 
there is a smooth action of ${\mathcal  G}$ on ${\Bbb E}$ which is
linear on the fibers and covers the action of ${\mathcal  G}$ on 
$\widetilde   Y$. Standard
constructions from linear algebra applied to any ${\mathcal  G}-$ locally 
homogeneous ${\Bbb E}$ give in natural way corresponding ${\mathcal  G}-$ 
locally homogeneous
vector bundles. In particular, all bundles $TY, {\Bbb C}\ell(Y),
{\rm End}\,{\Bbb E}\simeq {\Bbb E}^{*}\otimes {\Bbb E}$ are ${\mathcal  G}-$ 
locally
homogeneous \cite{mosc}. We shall require also that all constructions
associated with ${\mathcal  G}-$ locally homogeneous bundles to be 
${\mathcal  G}-$ 
equivariant. For example, for each $g$ in ${\mathcal  G}$ the commutative 
diagram is

$$
\begin{CD}
{\Bbb C}\ell(\widetilde   Y)@> >>{\rm End}\,{\widetilde  {\Bbb E}}\\
@V{ g}VV @VVgV\\
{\Bbb C}\ell(\widetilde   Y)@> >> {\rm End}\,{\widetilde  {\Bbb E}}
\end{CD}
\eqno{(C.1)}
$$
\vspace{0.3cm}

\noindent
Let ${\mathcal  O}$ denote a generalized Dirac operator associated to a locally
homogeneous bundle ${\Bbb E}$ over $Y$. We shall require ${\mathcal  G}-$ 
equivariance for ${\widetilde   \nabla}$ the lift of $\nabla$ to 
${\widetilde   {\Bbb E}}$ and therefore
the corresponding Dirac operator is then ${\mathcal  G}-$ invariant.

The main results can be stated as follows.

\medskip
\par \noindent
{\bf Theorem C.1} (Moscovici and Stanton [21]).\,\,\,{\em The following 
function can be defined, initially for $\Re(s^2)\gg 0$, by the formula

$$
{\rm log}Z(s,{\mathcal  O})\stackrel{def}{=}
\sum_{[\gamma]\in {\mathcal  E}_1(\Gamma)}
(-1)^q\frac{L(\gamma,{\mathcal  O})}
{|{\rm det}(I-P_h(\gamma))|^{1/2}}\frac{e^{-s\ell(\gamma)}}{m(\gamma)}
\mbox{,}
\eqno{(C.2)}
$$
where ${\mathcal  E}_1(\Gamma)$ is the set of those conjugacy classes 
$[\gamma]$
for which $Y_{\gamma}$ has the property that the Euclidean de Rham factor
of ${\widetilde   Y}_{\gamma}$ is 1-dimensional, the number $q$ is one-half the
dimension of the fiber of the center bundle $C(TY)$ over $Y_{\gamma}$, and 
$L(\gamma,{\mathcal  O})$ is the
Lefschetz number (see Ref. \cite{hott73-10}).
Furthermore ${\rm log}Z(s,{\mathcal  O})$ has a meromorphic continuation to
${\Bbb C}$ given by the identity

$$
{\rm log}Z(s,{\mathcal  O})={\rm log}{\rm det}^{'}
\left(\frac{{\mathcal  O}-is}{{\mathcal  O}+is}\right)
+\pi i\eta(s,{\mathcal  O})
\mbox{,}
\eqno{(C.3)}
$$
where $s\in i{\rm Spec}^{'}({\mathcal  O})$\,\,\,
$({\rm Spec}({\mathcal  O})-\{0\})$, and $Z(s,{\mathcal  O})$ satisfies the
functional equation

$$
Z(s,{\mathcal  O})Z(-s,{\mathcal  O})=e^{2\pi i\eta(s,{\mathcal  O})}
\mbox{.}
\eqno{(C.4)}
$$
}
\medskip

Suppose now $\chi: \Gamma\rightarrow U(F)$ be a unitary representation of 
$\Gamma$ on $F$. The Hermitian vector bundle 
${\Bbb F}={\widetilde   Y}\times_{\Gamma}F$ over $Y$ inherits a flat 
connection from the trivial connection on
${\widetilde   Y}\times F$. We specialize to the case of locally 
homogeneous Dirac operators 
${\mathcal  O}: C^{\infty}(Y,{\Bbb E})\rightarrow C^{\infty}(Y,{\Bbb E})$ in
order to construct a generalized operator ${\mathcal  O}_{\chi}$, acting on
spinors with coefficients in $\chi$. 
If ${\mathcal  O}: C^{\infty}(Y,V)\rightarrow C^{\infty}(Y,V)$ is
a differential operator acting on the sections of the vector bundle $V$,
then ${\mathcal  O}$ extends canonically to a differential operator 
${\mathcal  O}_{\chi}: C^{\infty}(Y,V\otimes{\Bbb F})\rightarrow 
C^{\infty}(Y,V\otimes{\Bbb F})$, uniquely characterized by the property that 
${\mathcal  O}_{\chi}$ is locally isomorphic to 
${\mathcal  O}\otimes...\otimes {\mathcal  O}$\,\,\,(${\rm dim}\,F$ times) 
\cite{mosc}.
 
One can repeat the arguments of the previous sections to construct a twisted
zeta function $Z(s,{\mathcal  O}_{\chi})$.

\medskip
\par \noindent
{\bf Theorem C.2} (Moscovici and Stanton [21]).\,\,\,{\em There exists a
zeta function  $Z(s,{\mathcal  O}_{\chi})$, meromorphic on ${\Bbb C}$, 
given for $\Re(s^2)\gg 0$ by the formula

$$
{\rm log}Z(s,{\mathcal  O}_{\chi})\stackrel{def}{=}
\sum_{[\gamma]\in {\mathcal  E}_1(\Gamma)}
(-1)^q{\rm Tr}\chi(\gamma)\frac{L(\gamma,{\mathcal  O})}
{|{\rm det}(I-P_h(\gamma))|^{1/2}}\frac{e^{-s\ell(\gamma)}}{m(\gamma)}
\mbox{;}
\eqno{(C.5)}
$$
moreover one has

$$
{\rm log}Z(0,{\mathcal  O}_{\chi})=\pi i\eta(0,{\mathcal  O}_{\chi})
\mbox{.}
\eqno{(C.6)}
$$
}

\end{document}